\documentclass[conference]{IEEEtran}
\IEEEoverridecommandlockouts
\usepackage{cite}
\usepackage{amsmath,amssymb,amsfonts}
\usepackage{algorithmic}
\usepackage{graphicx}
\usepackage{textcomp}
\usepackage{xcolor}
\usepackage{subcaption}
\usepackage{float}
\usepackage{comment}   

\newcommand{\commentout}[1]{}
\def\BibTeX{{\rm B\kern-.05em{\sc i\kern-.025em b}\kern-.08em
    T\kern-.1667em\lower.7ex\hbox{E}\kern-.125emX}}
\begin{document}

\title{Scaling Wideband Hybrid Beamforming for sub-THz Communication \\
}

\author{\IEEEauthorblockN{Amirmohammad Haddad, Oveys Delafrooz Noroozi, Canan Cebeci, Mark J. W. Rodwell and Upamanyu Madhow}
\IEEEauthorblockA{
University of California, Santa Barbara, CA, U.S.A}
\{amirmohammad, oveys, ccebeci, rodwell, madhow\}@ucsb.edu}


\maketitle

\begin{abstract}
We investigate the capacity attainable for a multiuser MIMO uplink as we scale both array size and bandwidth for regimes in which all-digital arrays incur excessive hardware complexity and power consumption. We consider a tiled hybrid beamforming architecture in which each tile, or subarray, is a phased array performing analog (or RF) beamforming, followed by DSP on the tile outputs.  For parameters compatible with sub-THz fixed access links, we discuss hardware and power consumption considerations for choosing tile size and the number of tiles.  Noting that the problem of optimal multiuser MIMO in our wideband regime is open even for the simplest possible channel models, we compare the spectral efficiencies attainable by a number of reasonable strategies for tile-level RF beamforming, assuming flexibility in the digital signal processing (DSP) of the tile outputs.  We consider a number of beam broadening approaches for addressing the ``beam squint'' incurred by RF beamforming in our wideband regime, along with strategies for sharing tiles among users.   Information-theoretic benchmarks are computed for an idealized MIMO-OFDM system, with linear per-subcarrier multiuser detection compared against an unconstrained complexity receiver.

\end{abstract}

\section{Introduction}
\label{Sec:Introduction}
\vspace{-0.1cm}
In this paper, we investigate fundamental limits of multiuser MIMO for sub-THz (100-150\,GHz) bands as we scale both array size and bandwidth.  Fully digital beamforming is challenging in this regime: recent work has proposed methods to reduce computational load of DSP in fully digital MU-MIMO \cite{cebeci2024scaling, noroozi2025scaling}, but the complexity and power consumption of downconversion stages and high-speed data converters pose fundamental constraints.  We therefore focus on the tradeoff between hardware architecture and system performance for tiled hybrid beamforming as we scale the number of array elements while targeting large fractional bandwidth of $\sim20$\% (e.g., about 30 GHz bandwidth at a carrier frequency of 140 GHz). While much of the literature on hybrid beamforming focuses on transmit precoding (e.g., \cite{gao2021wideband, dai2022delay}), we focus here on receive beamforming: utilizing reciprocity is key for scaling channel estimation with array size and bandwidth, hence receive beamforming is a pre-requisite for transmit beamforming.  We seek insight for the simplest possible model, a static LoS channel for each user and an $N$-element linear array at half-wavelength spacing.  While the model is simple, it is well matched to applications such as fixed wireless access.

Each tile is a phased array with $N_a$ elements performing analog, or RF, beamforming, and there are $N_d$ such tiles.  For fixed array size $N = N_a N_d$, there is an inherent tension in how the array should be organized.  Increasing the number of elements $N_a$ per tile improves power efficiency, but with a corresponding reduction in flexibility of processing due to the reduction in the number of independent digital streams $N_d$. This translates to reduced capacity due to increased vulnerability to beam squint and the smaller number of users that can be supported.  Optimal MU-MIMO processing is an open problem even in the simple setting we consider: while classic MU-MIMO techniques can be applied after per-tile RF beamforming, the choice of RF beamformers for optimizing system performance is not known.  We therefore compare a number of sensible strategies, including adaptations of prior work in the literature, via information-theoretic benchmarks.

\noindent
{\bf Contributions and Organization:} We discuss hardware design considerations on tile size and number of tiles for sub-THz arrays in Section \ref{Sec:Hardware}. Section \ref{Sec:SystemModel} describes the wideband MU-MIMO system model and formulates the information-theoretic benchmarks to be used in our evaluation. Section \ref{Sec:Broadbeams} describes RF beam broadening for addressing beam squint.  Section \ref{Sec:ResultsDiscussion} compares a number of beamforming strategies for single-user and multiuser settings.  An encouraging conclusion is that simple strategies involving phase-only RF beamforming incur little performance loss relative to, and often outperform, more complex strategies requiring both amplitude and phase control.

\begin{figure}[H]
    \centering
    \includegraphics[width=0.8\linewidth, height=0.45\linewidth]{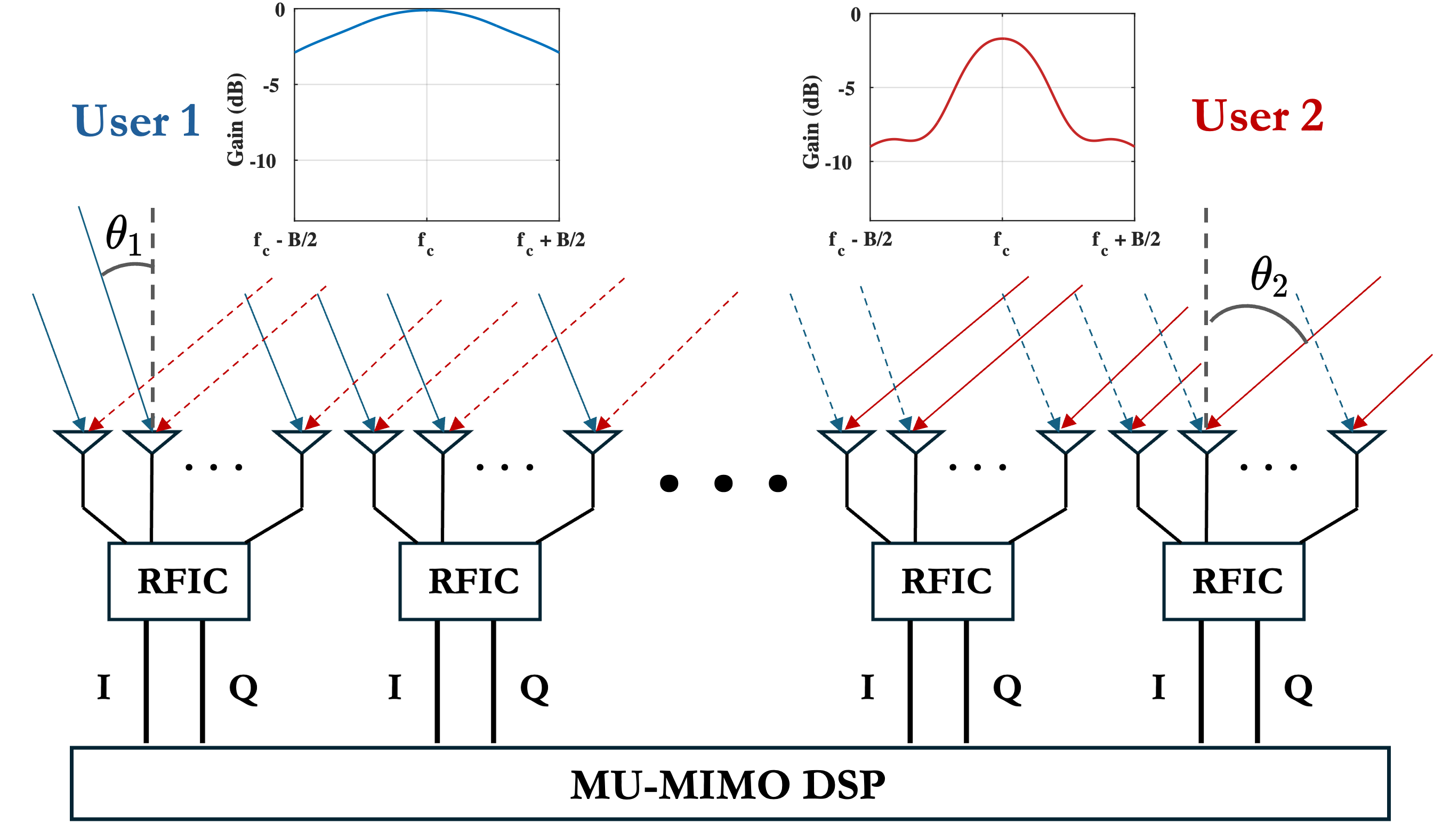}
    \caption{\small{Wideband MU-MIMO communication via a tiled antenna array}}
    \label{SubFig:ProbStatement}
\end{figure}

\section{Hardware-Aware Design Insights}
\label{Sec:Hardware}
\begin{figure}[h]
    \begin{subfigure}{0.5\textwidth}
        \centering
        \includegraphics[width=0.7\linewidth, height=0.5\linewidth]{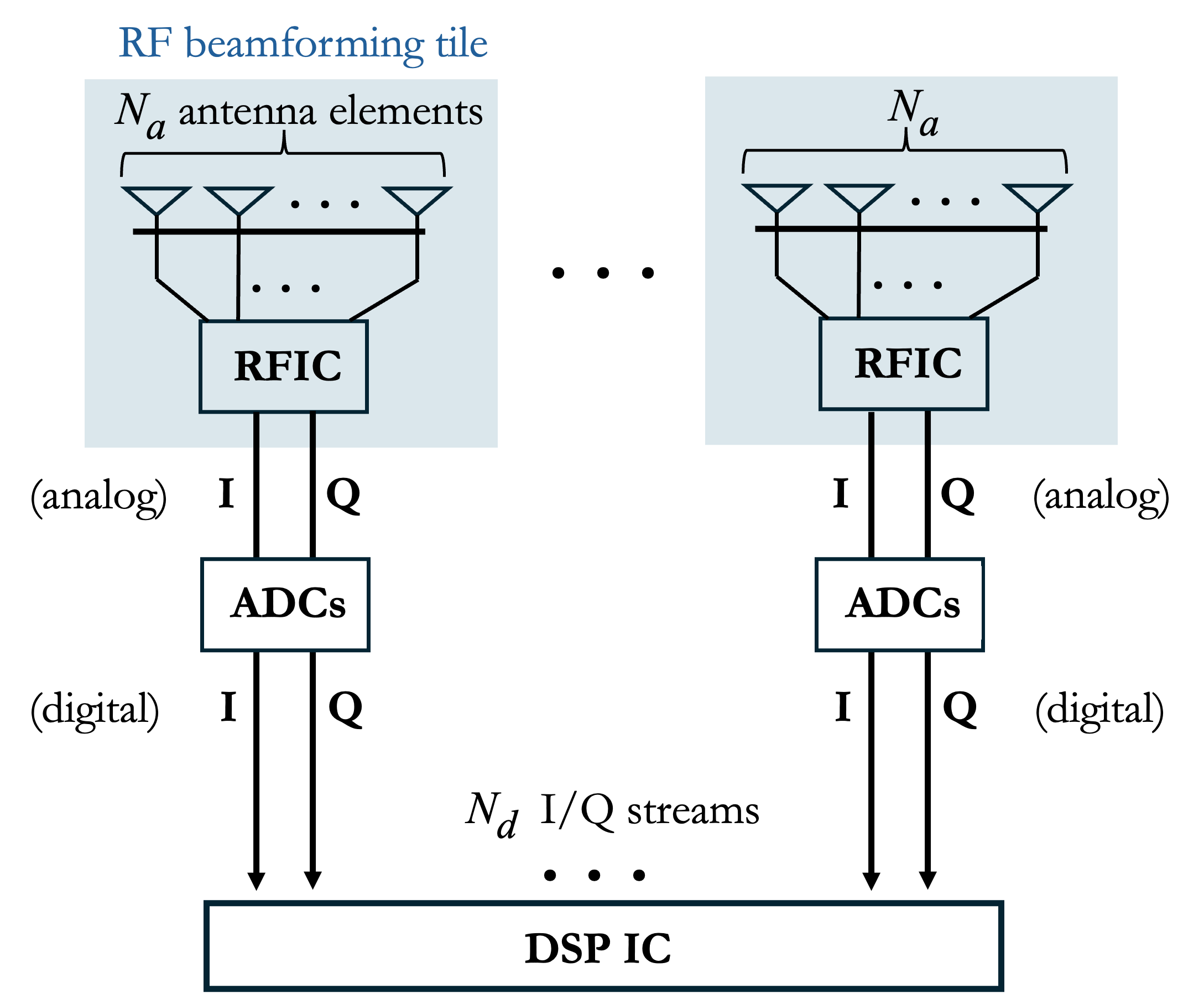}
        \caption{\small{Hybrid beamforming schematic}}
        \label{SubFig:HybridSys}
    \end{subfigure}
    \begin{subfigure}{0.5\textwidth}
        \centering
        \includegraphics[width=0.7\linewidth, height=0.3\linewidth]{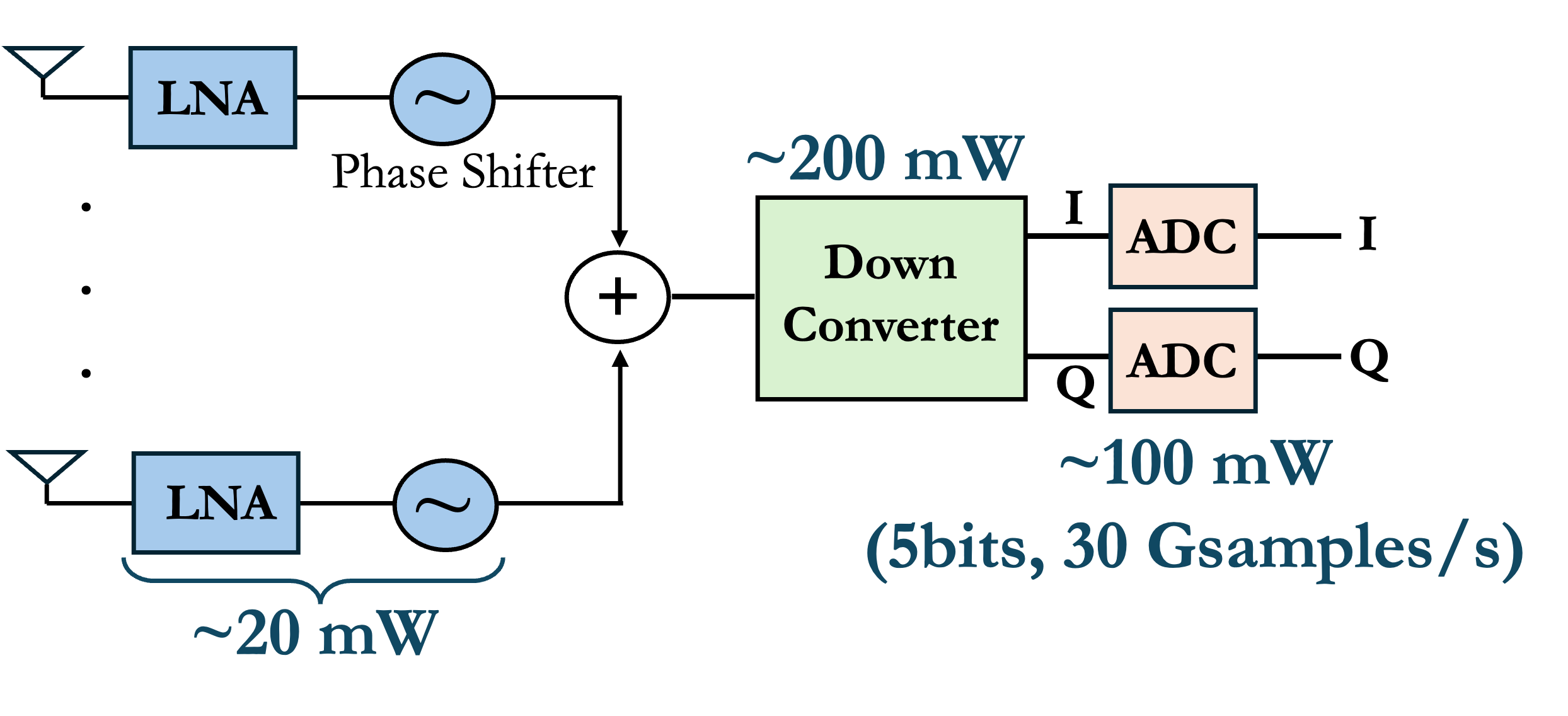}
        \caption{\small{Per-element and per-tile hardware components}}
        \label{SubFig:PerTileSys}
    \end{subfigure}    

    \caption{\small Hybrid array hardware structure and power components.
Subfigure~(a) shows the $N_d$ RF tiles, each with $N_a$ antennas, while
subfigure~(b) highlights the main per-element and per-tile hardware
blocks with typical power values.
}
    \label{Fig:SystemModel}
\end{figure}

We now discuss hardware considerations in designing hybrid beamforming architectures at a carrier frequency of 140 GHz and bandwidth of 28 GHz.  We consider a $N=256$-element half-wavelength spaced array, which spans a little over 25 cm, and is compatible, for example, with compact base stations or access points installed on lampposts or utility poles for providing last hop fixed wireless access in a neighborhood.  We first discuss power consumption considerations in choosing the tile size $N_a$ and number of tiles $N_d$.  Figure \ref{SubFig:PerTileSys} displays typical numbers for key RF and mixed signal components.  The power consumption for the low noise amplifier (LNA) and phase shifter required for each element is in the 10s of mW.  The power consumption for an RF chain per tile includes the mixer used for downconversion, and the ADCs for the in-phase (I) and quadrature (Q) components.  We roughly estimate these as 200 mW and 100 mW per ADC respectively based on the state of the art in the research literature. The specific numbers are less important than the fact that these are an order of magnitude higher than the power consumption for the LNA and phase shifter, which motivates reducing the number of RF chains, and hence the number of tiles.  Based on these rough estimates,  we estimate the total power consumption prior to DSP as 
\begin{equation} \label{Ptot}
 P_{\mathrm{tot}} \approx 20N + 400N_d \ \text{mW}  \nonumber
\end{equation}
corresponding to a per-element power consumption of $P_{\mathrm elt} = 20 + 400/N_a = 20 + 400 N_d/N \ \text{mW}$.  

Based on the preceding estimates, pure RF beamforming ($N_a=N=256$, $N_d=1$) minimizes power consumption, but of course, this choice limits multiuser capability.  In addition, the analog tile size $N_a$ is limited by wire-length-induced loss, estimated at about $0.2$ dB/mm at 140 GHz. Since the wavelength is about 2 mm, the maximum RF trace length within a tile grows approximately as $N_a/2$ mm at 140 GHz, assuming that the RFIC controlling the tile is placed to minimize this length.  Thus, if we limit the insertion loss within a tile to about 2-3 dB, candidate tile sizes lie in the range $N_a = 16-32$. With a loss of roughly $0.2$ dB/mm at the same frequency, limiting the insertion loss within a tile to about $2-3$ dB leads us to consider candidate analog tile sizes $N_a=16$ and $N_a=32$. Plugging into our coarse power consumption estimates in (\ref{Ptot}), we obtain 11.5 W for $N_a=16$ (hence $N_d =16$) and 8.3 W for $N_a=32$ (hence $N_d =8$).  We focus on the latter setting in our numerical examples, while noting that it presents a greater challenge in terms of system performance: beam squint is worse, and the number of simultaneous users that can be supported is smaller.

\section{System Model and Performance Metrics}
\label{Sec:SystemModel}

We consider a uniform linear array (ULA) partitioned into $N_d$ tiles, each containing $N_a$ antenna elements with inter-element spacing
$\lambda_c/2$. For user $k$ with angle $\theta_k$, the frequency-dependent spatial frequency is
\begin{equation}
\Omega_k(f)
=
\frac{f}{f_c}\Omega_{c,k}
=
\frac{f}{f_c}\pi\sin\theta_k ,
\label{eq:Omega}
\end{equation}
where $\Omega_{c,k}=\pi\sin\theta_k$ is the center spatial frequency. The steering vector for an $N$-element ULA at spatial frequency $\Omega$ is
\begin{equation}
a_N(\Omega)
=
\begin{bmatrix}
1 &
e^{j\Omega} &
\cdots &
e^{j(N-1)\Omega}
\end{bmatrix}^{T}.
\label{eq:steer}
\end{equation}

For the $m$-th tile and user $k$, the per-tile steering response is
\begin{equation}
\Phi_{mk}(f)
=
a_{N_a}(\Omega_k(f))
\,
e^{j(m-1)N_a\Omega_k(f)}
\in\mathbb{C}^{N_a\times 1}.
\end{equation}
The channel between user $k$ and tile $m$ is modeled as
\begin{equation}
h_{mk}(f)
=
\alpha_{mk}\,\Phi_{mk}(f)
\in\mathbb{C}^{N_a\times 1},
\label{eq:hmk}
\end{equation}
where $\alpha_{mk}$ is the complex gain of that link.

\vspace{2mm}
Following tile-level RF beamforming, our nominal MU-MIMO DSP is per-subcarrier linear minimum mean squared error (LMMSE) followed by single-user decoding. We compare its performance against an unconstrained-complexity benchmark.  Tile $m$ applies an analog beamforming vector $w_m \in \mathbb{C}^{N_a}$, producing the following effective channel for user $k$ at frequency $f$:
\begin{equation} \label{rf_beamformer_out}
    h^{(\mathrm{eff})}_{mk}(f)
    =
    \frac{w_m^{H}}{\|w_m\|_2} h_{mk}(f)
    \in\mathbb{C}.
\end{equation}
Stacking across all tiles yields the effective channel vector for user $k$:
\begin{equation}
h_k(f)
=
\begin{bmatrix}
h^{(\mathrm{eff})}_{1k}(f) &
h^{(\mathrm{eff})}_{2k}(f) &
\cdots &
h^{(\mathrm{eff})}_{N_d k}(f)
\end{bmatrix}^{T}
\in\mathbb{C}^{N_d\times 1}.
\label{eq:hk} \nonumber
\end{equation}
Collecting all users gives the wideband effective channel matrix
\begin{equation}
\mathbf{H}(f)
=
\big[\, h_1(f)\; h_2(f)\; \cdots\; h_K(f)\,\big]
\in\mathbb{C}^{N_d\times K}.
\label{eq:Hf}
\end{equation}
Digital MU–MIMO processing is performed independently on each subcarrier. For user $k$, let $\tilde{\mathbf{H}}_k(f)$ denote $\mathbf{H}(f)$ with the $k$-th column removed. The interference-plus-noise covariance is given by
\begin{equation}
    \mathbf{R}_k(f)
    =
    \tilde{\mathbf{H}}_k(f)\tilde{\mathbf{H}}_k^{H}(f)
    +
    2\sigma^2 I,
\label{eq:Rk}
\end{equation}
where $\sigma^2$, the noise variance per dimension, is unchanged because of the normalization in (\ref{rf_beamformer_out}).
The SINR for user $k$ using per-subcarrier LMMSE is
\begin{equation}
\mathrm{SINR}_k(f)
=
h_k^{H}(f)\,
 \mathbf{R}_k^{-1}(f)\,
h_k(f).
\label{eq:SINR}
\end{equation}
The corresponding spectral efficiency, averaging across the system bandwidth, is 
\begin{equation}
r_k
=
\frac{1}{B}
\int_{-B/2}^{B/2}
\log_2\!\big(1+\mathrm{SINR}_k(f)\big)\,
df .
\label{eq:rate}
\end{equation}
with SINR replaced by SNR for a single user.

The preceding benchmarks are based on linear MU-MIMO processing, whose suboptimality can be quantified by comparing the sum rate against that attained by unconstrained MU-MIMO processing, given by 
\begin{equation}
    \hspace{-0.25cm}
    r_s  = \sum_{k=1}^{K} r_k \;\le\;
    \frac{1}{B} \int_{-B/2}^{B/2}
    \log_2 \det\!\left(
        \mathbf{I} + \frac{\mathbf{H}^H(f)\mathbf{H}(f)}{2\sigma^2}
    \right) df. \nonumber
\end{equation}



\section{Beam Squint Problem and Broad Beam Design}
\label{Sec:Broadbeams}
For wideband systems, we see from (\ref{eq:Omega}) that the spatial frequency can vary significantly across the band for steeper angles of arrival. For a fractional bandwidth $\beta = B/f_c$, the variation in spatial frequency over the band is $\Delta\Omega=|\Omega_c|\,\beta$. Thus, a standard narrow beam (i.e., a spatial matched filter to the response at the carrier frequency) does not remain aligned at other frequencies.  In order to address this phenomenon of beam squint, we consider broadening the beam produced by the tile-level RF beamformers.


A practical and well-known approach for synthesizing broad beams is to impose a
quadratic phase progression across the elements, as demonstrated for both linear~\cite{fontenauQuadratic,sayidmarie2013synthesis} and planar arrays ~\cite{zhang2025beam,petersson2022energy}. Similarly, we consider the following quadratic phase definition across the elements 
\begin{equation}
    \phi[n] = \Omega_c n + \frac{\Delta\Omega}{2N} n^2, \qquad n=-\tfrac{N}{2},\ldots,\tfrac{N}{2},
\end{equation}
where $\Delta\Omega$ governs the beamwidth. The corresponding weight vector for tile $m$ is
\begin{equation}\label{quadraticWeights}
    w_m = [\,w_{m1},\ldots,w_{mN_a}\,]^T,
    \quad
    w_{mi} = e^{j\,\phi[i-(N/2+1)]}.
\end{equation}
To obtain insight into why the quadratic term broadens the beam, let us treat the element index $n$ as a continuous variable. We see that the instantaneous spatial frequency varies across the array as follows:
\begin{equation}
    \frac{d\phi}{dn}
    = \Omega_c + \frac{\Delta\Omega}{N}n
    \in \left[\Omega_c - \frac{\Delta\Omega}{2},
              \, \Omega_c + \frac{\Delta\Omega}{2}\right].
\end{equation}
Broad beams synthesized in this fashion play an important role in the hybrid beamforming strategies that we discuss next.

\section{Results and Discussion}
\label{Sec:ResultsDiscussion}
Informed by the hardware considerations above, we consider a hybrid architecture with $N_d\!=\!8$ tiles and $N_a\!=\!32$ antenna elements per tile.
Under wideband operation with $\beta = 0.2$, we compare the spectral efficiency of different cases to obtain the most suitable beamforming strategies.



\subsection{Single-user performance}
\begin{figure}[h]
    \centering
    \begin{subfigure}[t]{\linewidth}
        \centering
        \includegraphics[width=\linewidth]{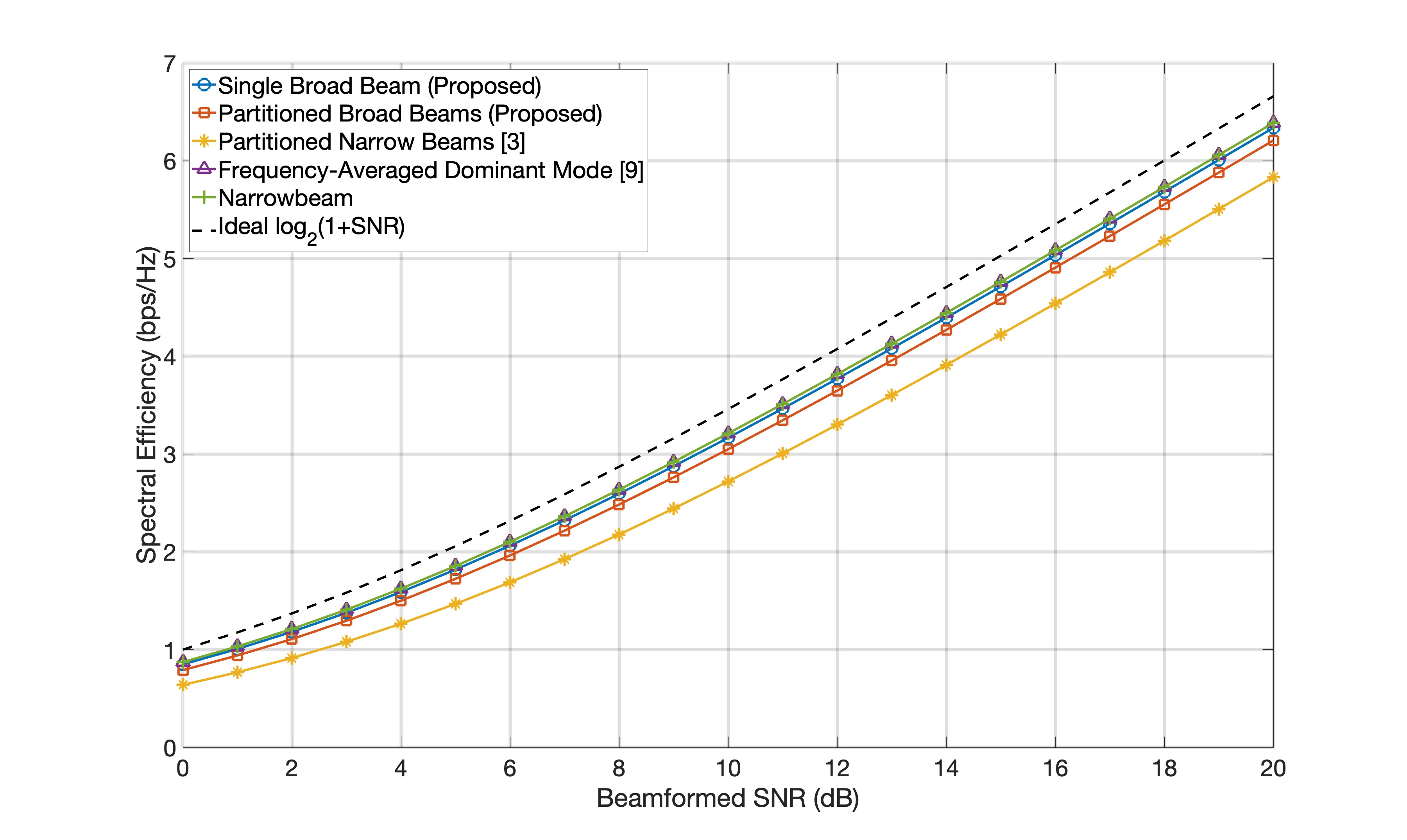}
        \caption{\small Single-user capacity at $\theta = 15^{\circ}$}
        \label{fig:Single15}
    \end{subfigure}
    \begin{subfigure}[t]{\linewidth}
        \centering
        \includegraphics[width=\linewidth]{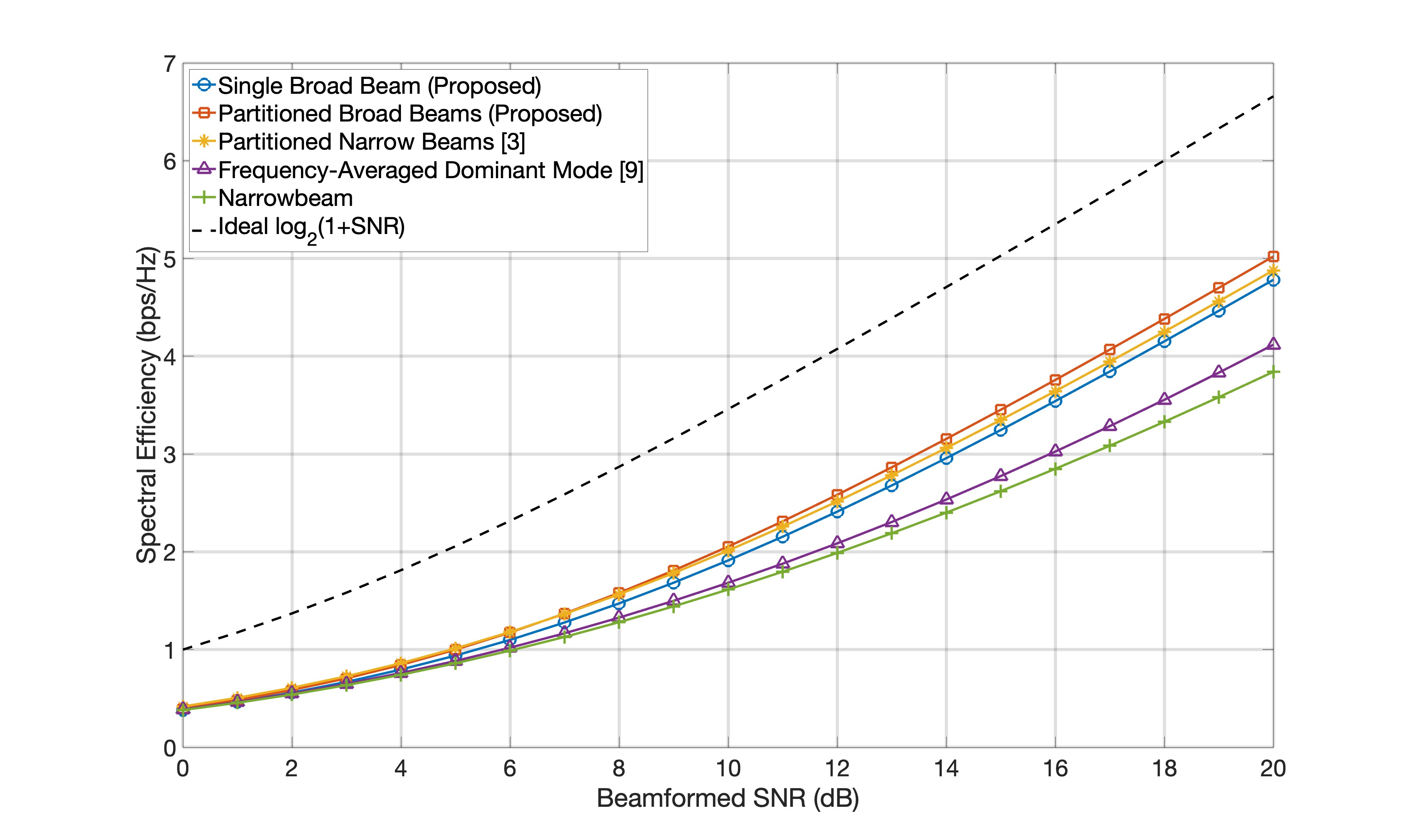}
        \caption{\small Single-user capacity at $\theta = 55^{\circ}$}
        \label{fig:Single55}
    \end{subfigure}
    \caption{\small Single-user rate for different analog beamforming
    strategies and ideal $\log_2(1+\mathrm{SNR})$ limit at steering angles (a) $\theta = 15^\circ$ and (b) $\theta = 55^\circ$.}
    \label{fig:SingleUser}
\end{figure}

In the single-user setting, analog beamforming is sufficient, and we compare
four wideband strategies against a narrowband baseline and the ideal
$\log_2(1+\mathrm{SNR})$ limit.
\begin{itemize}
    \item \textbf{Single broad beam:} All tiles use a broad beam covering the entire range of spatial frequencies across the band, as in (\ref{quadraticWeights}).
    \item \textbf{Partitioned broad beams:} Each of the $N_d$ tiles forms a broad beam covering a fraction $\frac{1}{N_d}$ of the spatial frequency spread across the band.
    \item \textbf{Partitioned narrow beams~\cite{gao2021wideband}:} A beam-partitioning method that combines multiple narrowband beams, each sized using its 3\,dB beamwidth, to cover the range of spatial frequencies.
    \item \textbf{Frequency-averaged dominant mode, adapted from~\cite{sohrabi2017hybrid}:} The beamforming weight vector is obtained as the eigenvector corresponding to the largest eigenvalue of the frequency-averaged channel covariance matrix, with each element normalized to unit magnitude so as to require phase-only RF beamforming.
 
\end{itemize}
Fig.~\ref{fig:SingleUser} compares the single-user rate, obtained by replacing SINR by SNR in \eqref{eq:rate}, for the described wideband RF beamforming strategies at two steering angles. At
an angle close to broadside (Fig.~\ref{fig:Single15}), the narrowband beamformer provides the highest rate, with the loss from beam broadening exceeding the distortion due beam squint. Among the wideband methods, \cite{sohrabi2017hybrid} remains the closest to the narrowband response. The single broad beam performs better than the partitioned-beam designs, which incur extra loss due to the nulls between adjacent beams. Between the two partitioned strategies, the proposed partitioned broad beam achieves slightly higher rates than the narrowbeam-based
design of \cite{gao2021wideband}. At a steep angle (Fig.~\ref{fig:Single55}), beam squint dominates and all
methods fall noticeably below the ideal limit, with a substantially larger gap than in Fig.~\ref{fig:Single15}. The method of
\cite{sohrabi2017hybrid} remains only slightly better than the narrowband
design. Both partitioned-beam approaches outperform the single broad beam
because the latter must form a much wider pattern, leading to greater gain
loss. In this regime as well, the proposed partitioned design yields a
consistent but modest improvement over \cite{gao2021wideband}. Since the partitioned beams offer only a negligible benefit, we adopt the single broad beam in the remainder of the paper for its simplicity and ease of
implementation.

\subsection{Multiuser MIMO}
\subsubsection{Two-User Hybrid MU--MIMO}
\begin{figure}[h]
    \centering
    \begin{subfigure}[t]{\linewidth}
        \centering
        \includegraphics[width=\linewidth]{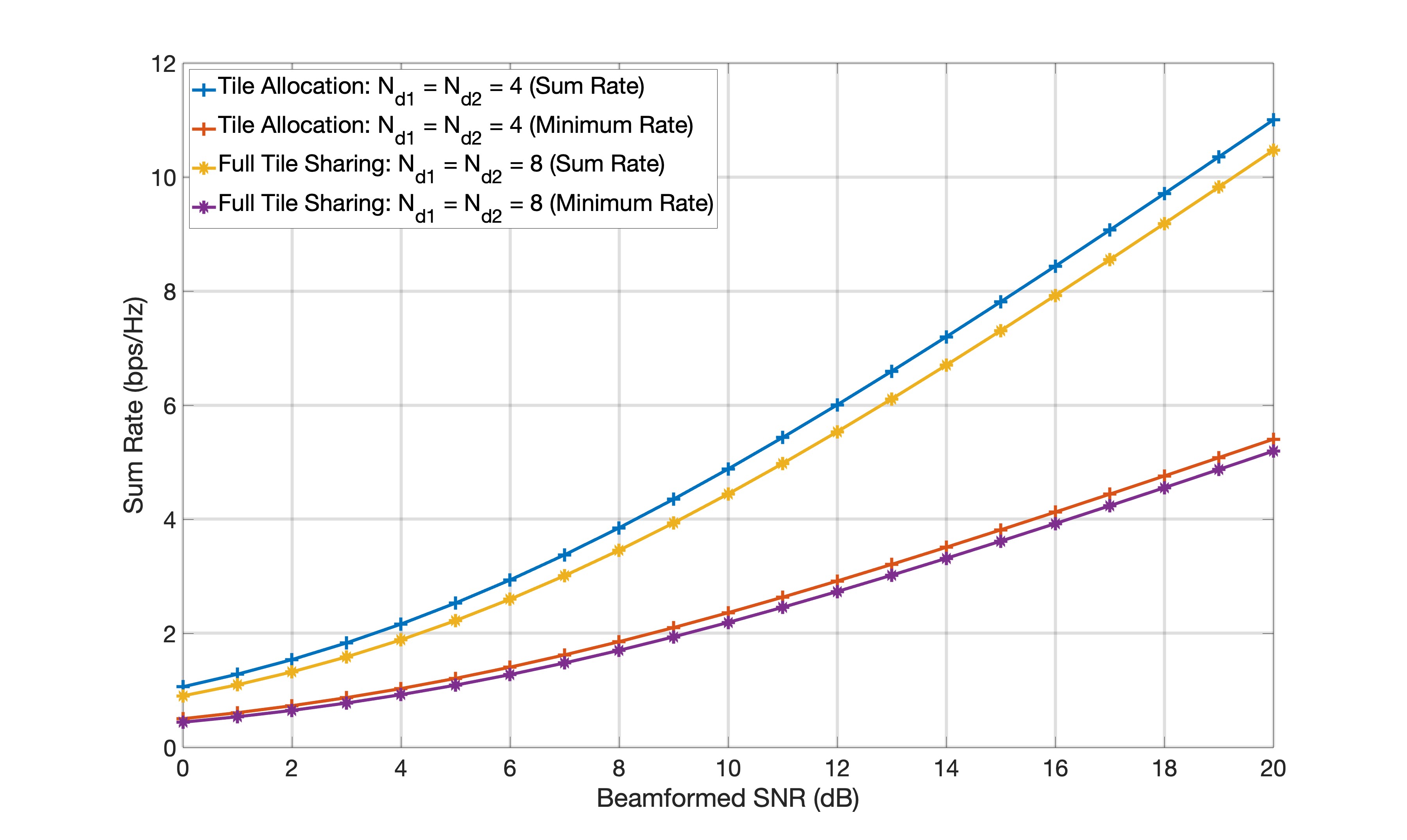}
        \caption{\small Two users at $(\theta_1, \theta_2) = \left(10^\circ, 15^{\circ}\right)$}
        \label{fig:TwoUser10_15}
    \end{subfigure}
    \begin{subfigure}[t]{\linewidth}
        \centering
        \includegraphics[width=\linewidth]{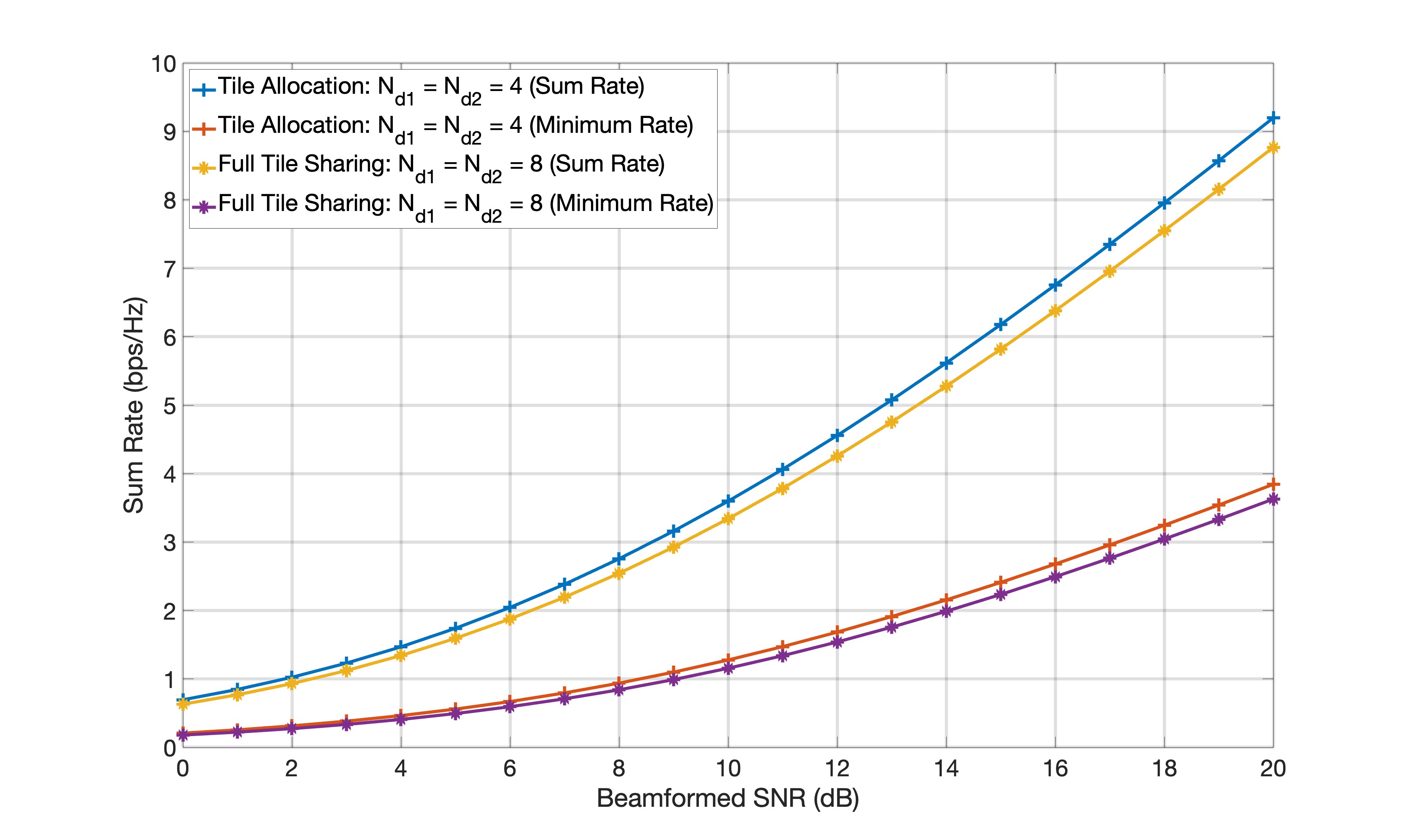}
        \caption{\small Two users at $(\theta_1, \theta_2) = \left(15^\circ, 55^{\circ}\right)$}
        \label{fig:TwoUser15_55}
    \end{subfigure}
    \begin{subfigure}[t]{\linewidth}
        \centering
        \includegraphics[width=\linewidth]{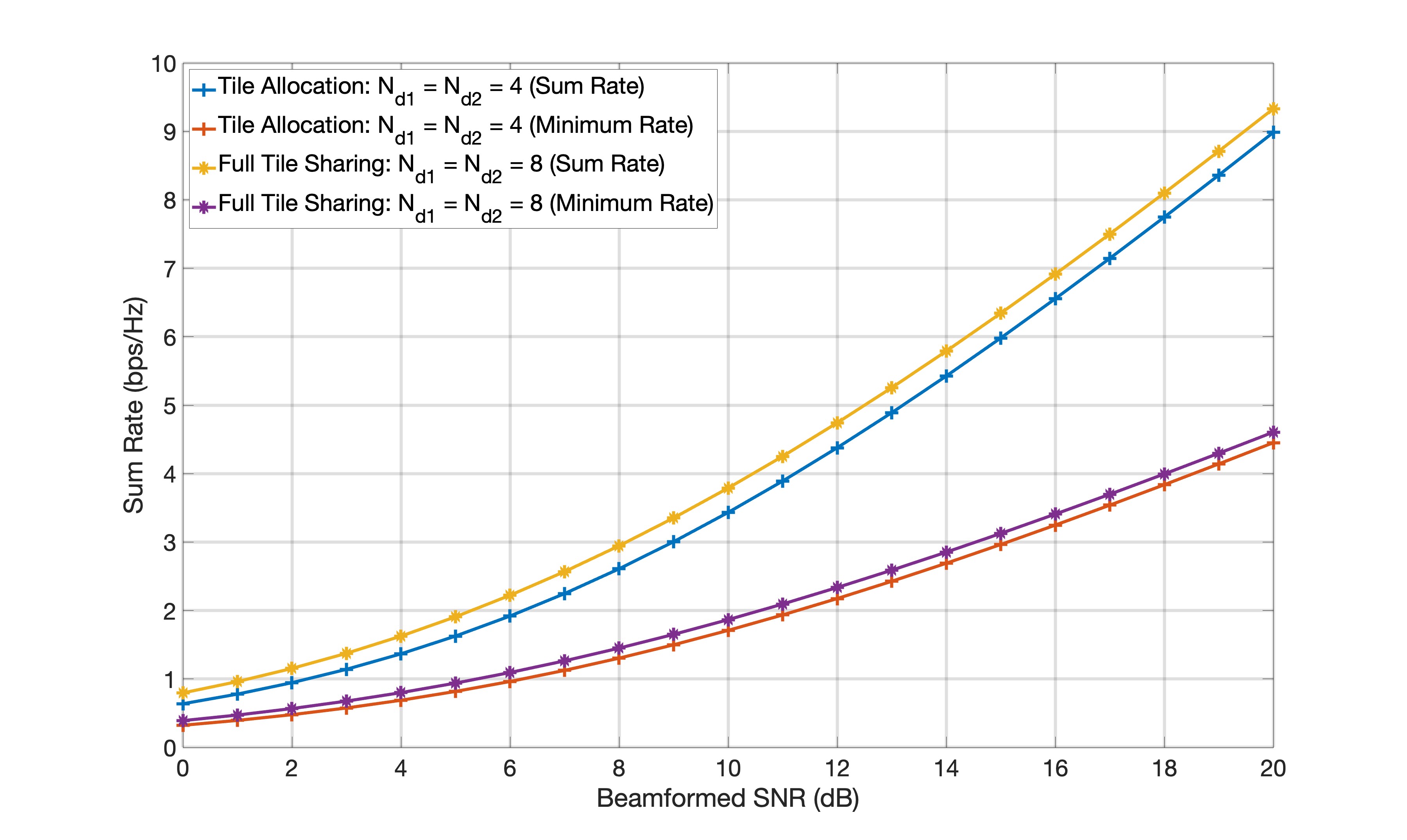}
        \caption{\small Two users at $(\theta_1, \theta_2) = \left(50^\circ, 55^{\circ}\right)$}
        \label{fig:TwoUser50_55}
    \end{subfigure}
    \caption{\small Sum rate performance for two-user hybrid MU--MIMO under two analog beamforming strategies: disjoint tile allocation (each user assigned four tiles) and full tile sharing (both users utilizing all eight tiles). Results are shown for user angle pairs (a) $(\theta_1,\theta_2)=(10^\circ,15^\circ)$, (b) $(15^\circ,55^\circ)$, and (c) $(50^\circ,55^\circ)$.
}
    \label{fig:TwoUser}
\end{figure}

\begin{figure}[h]
    \centering
    \includegraphics[width=0.95\linewidth, height = 0.5\linewidth]{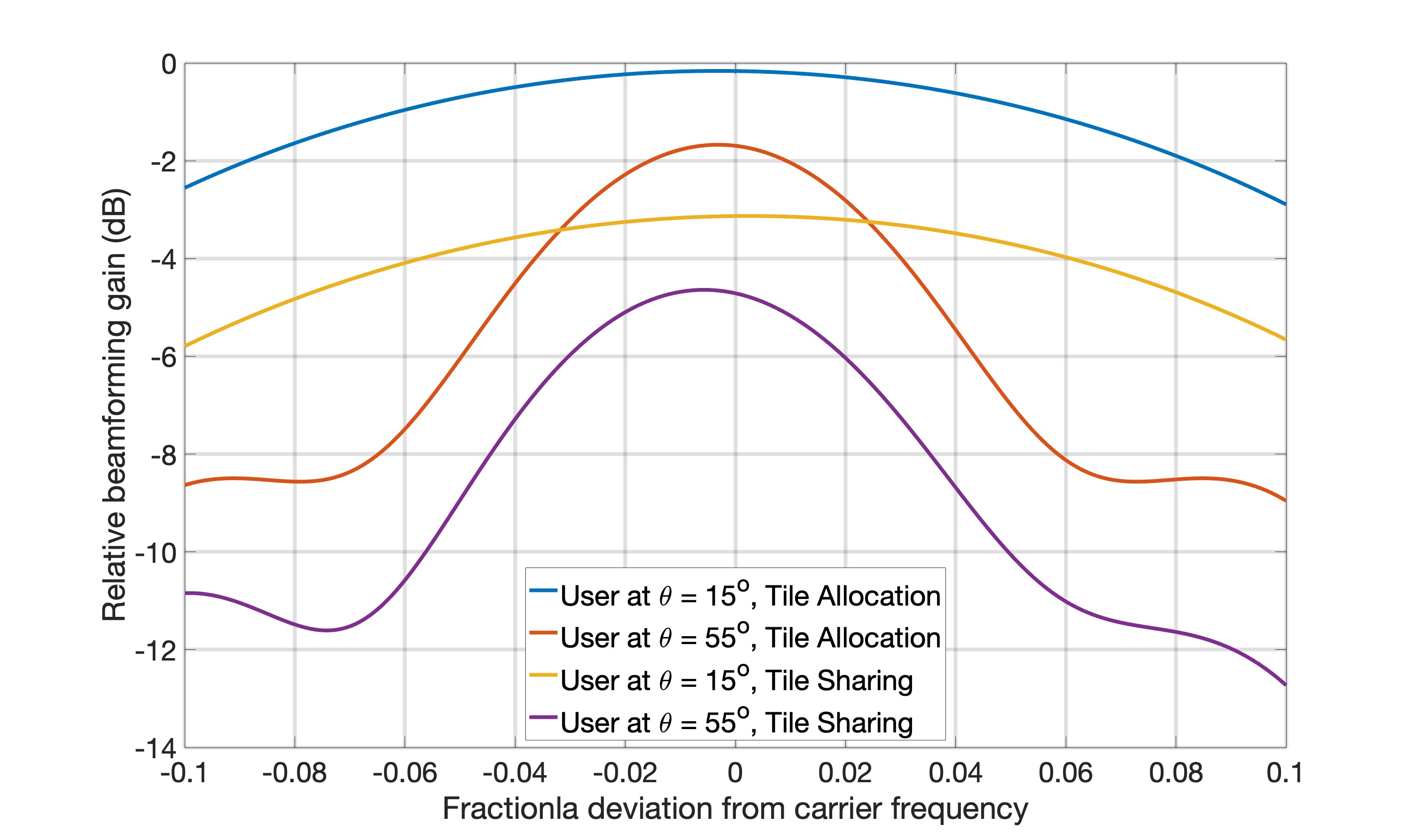}
    \caption{\small Comparison of beam patterns produced by tile allocation and
    tile sharing for a two-user scenario at $(\theta_1,\theta_2) = (15^\circ,55^\circ)$.}

    \label{fig:BeamPattern}
\end{figure}

We next examine tile-allocation strategies in a two-user MU--MIMO setting. 
Fig.~\ref{fig:TwoUser} presents a comparison between disjoint tile allocation and full tile sharing based on the resulting sum rate and the minimum-rate user (i.e., the user with the steeper AoA). In most cases, allocating separate tiles to each user yields a higher sum rate. While there are some settings in which tile sharing performs slightly better, the benefit is outweighed by the fact that it requires amplitude as well as phase control. The beam patterns in Fig.~\ref{fig:BeamPattern} provide further insight, showing that tile sharing experiences noticeably larger gain degradation across frequency, which explains its reduced performance compared to disjoint tile allocation.
Figs.~\ref{fig:TwoUser}
further show that the digital beamformer can effectively separate users even 
when their angles are close. 

\subsubsection{The impact of MU-MIMO DSP}

In hybrid arrays, the number of tiles (or RF chains) determines the maximum number of users that can be spatially separated. In order to understand the performance benefits from the additional computational complexity and energy consumption incurred by digital MU-MIMO DSP, we introduce a simple benchmark: assign one tile to each user and rely solely on RF beamforming followed by single-user DSP. 

\begin{figure}[h]
    \centering
    \begin{subfigure}[t]{\linewidth}
        \centering
        \includegraphics[width=\linewidth]{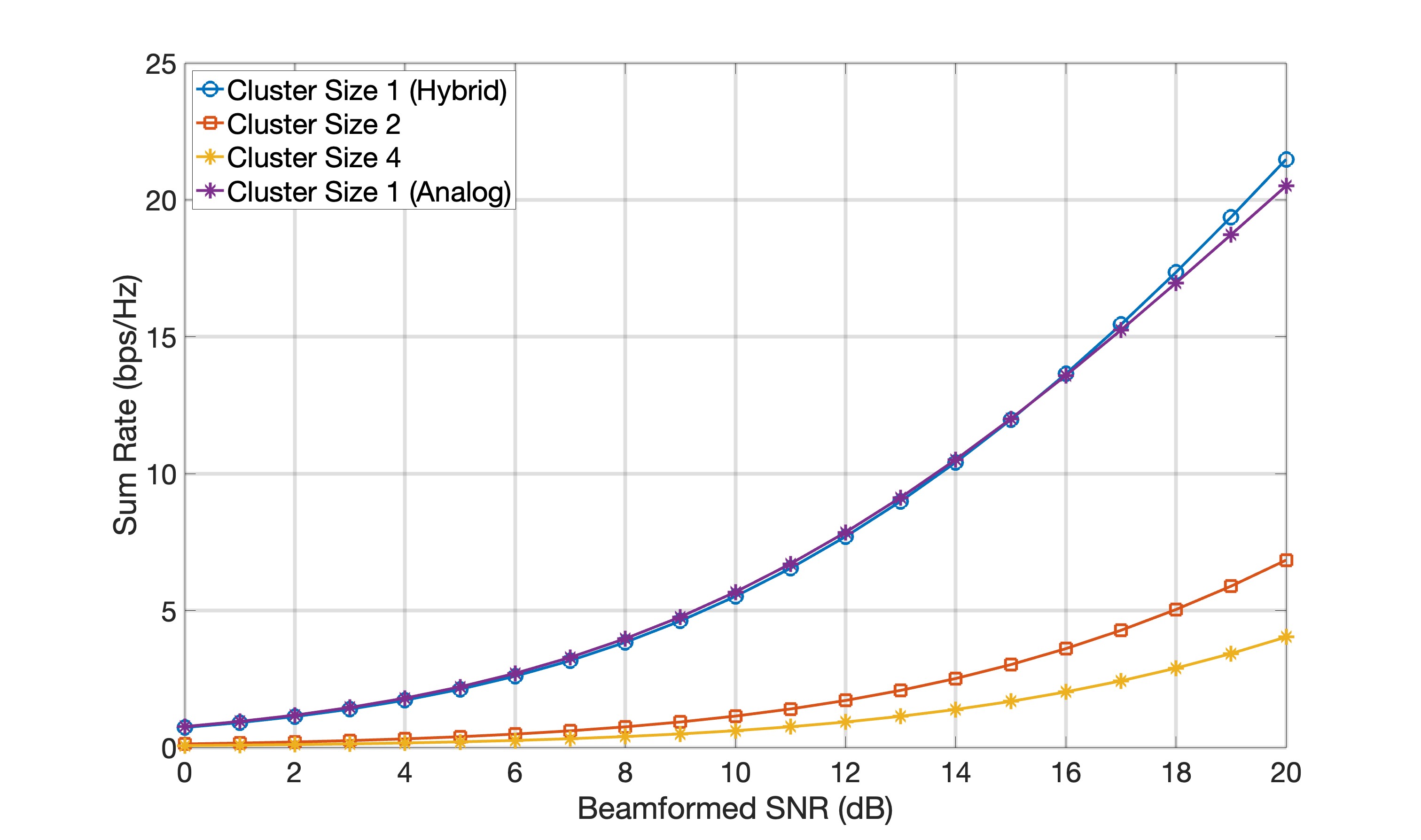}
        \caption{\centering \small
        Separated user configuration with angles
        \scriptsize $\{\theta_1,\theta_2,\ldots,\theta_8\} = \{-55^\circ,-40^\circ,-25^\circ,-10^\circ,10^\circ,25^\circ,40^\circ,55^\circ\}$}
        \label{fig:SepUsers}
    \end{subfigure}

    \begin{subfigure}[t]{\linewidth}
        \centering
        \includegraphics[width=\linewidth]{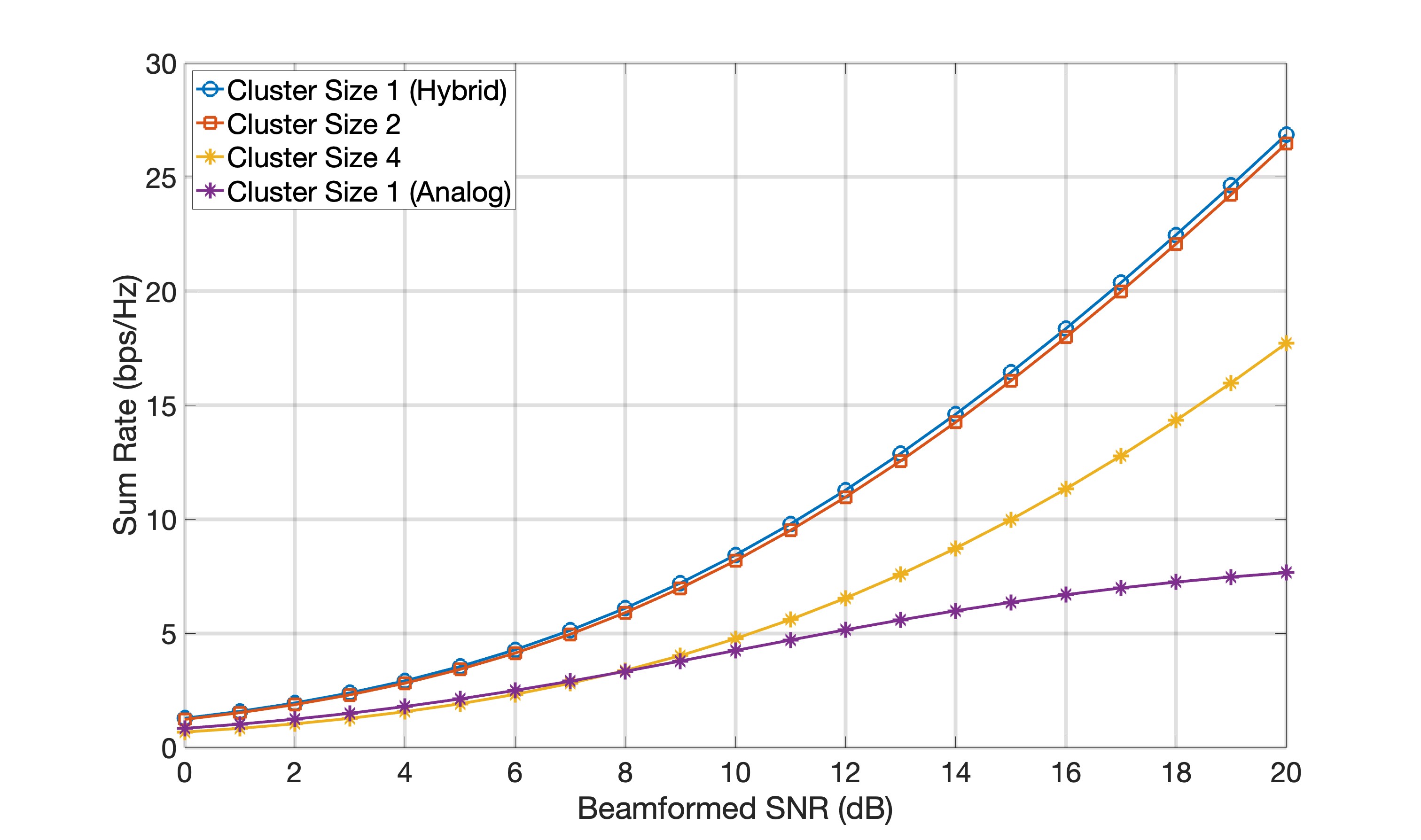}
        \caption{\centering \small Close user configuration with angles
        \scriptsize $\{\theta_1,\theta_2,\ldots,\theta_8\} = \{1^\circ,3^\circ,5^\circ,7^\circ,49^\circ,51^\circ,53^\circ,55^\circ\}$}
        \label{fig:CloseUsers}
    \end{subfigure}
    \caption{\small
    Angle configurations used to evaluate multiuser hybrid beamforming performance.  
    Let the user angles be denoted $\{\theta_1,\theta_2,\ldots,\theta_8\}$. We consider three clustering structures:\\
    Cluster size 1: $\{\theta_1\},\{\theta_2\},\ldots,\{\theta_8\}$\\
    Cluster size 2: $\{\theta_1,\theta_2\},\{\theta_3,\theta_4\},\{\theta_5,\theta_6\},\{\theta_7,\theta_8\}$\\
    Cluster size 4: $\{\theta_1,\ldots,\theta_4\},\{\theta_5,\ldots,\theta_8\}$.\\
    Subfigure (a) uses widely separated users, while subfigure (b) uses two groups of nearby users.
    }

    \label{fig:MultiUser}
\end{figure}

To understand the impact of digital MU-MIMO processing, we consider two arrangements of eight users, equal to the number of tiles, with one configuration having well-separated angles and the other containing
two groups of nearby users. Fig.~\ref{fig:MultiUser} compares the sum-rate performance of different clustering and tile-assignment strategies under these two scenarios. A cluster size $k$ indicates that the users are partitioned into clusters of $k$ users, with each cluster assigned $k$ tiles. All tiles within a given cluster use the same RF beamforming coefficients, designed to produce a single broad beam whose beamwidth is wide enough to cover the union of the spatial frequency ranges of all users in that cluster.

As shown in Fig.~\ref{fig:SepUsers}, when users are sufficiently separated, the interference is negligible, and RF beamforming (with single-user DSP) performs close to hybrid beamforming. 
In contrast, Fig.~\ref{fig:CloseUsers} illustrates that when users are closely spaced and interference becomes significant, MU-MIMO DSP for interference suppression provides significant benefits.

\section{Conclusion}
\label{Sec:conclusion}
Our work shows that hybrid beamforming with phase-only RF beamformers can effectively support wideband multiuser MIMO.  For a single-user, a properly designed broad beam maintains stable gain across the band, making elaborate analog control unnecessary for addressing beam squint.  For multiple users, this approach to per-tile beamforming, along with simple tile allocation strategies, work as well as more complex approaches. In order to assess the impact of MU-MIMO DSP after tile-level beamforming, it was compared against RF beamforming for each user with single-user DSP.  We find that the latter strategy is quite effective when users are well separated, whereas digital MU-MIMO processing based on tile-level beamforming outputs provides substantial performance gains for closely spaced users.  These results emphasize the benefits of interference-aware design for analog and digital processing to address complexity-performance and power–efficiency tradeoffs in wideband hybrid beamforming.

\section*{Acknowledgments }
This work was supported by the Center for Ubiquitous Connectivity (CUbiC) through the SRC–DARPA JUMP 2.0 program, and by the National Science Foundation under grant CNS-21633.


\bibliographystyle{IEEEtran}
\bibliography{Ref}

\begin{thebibliography}{1}
\providecommand{\url}[1]{#1}
\csname url@samestyle\endcsname
\providecommand{\newblock}{\relax}
\providecommand{\bibinfo}[2]{#2}
\providecommand{\BIBentrySTDinterwordspacing}{\spaceskip=0pt\relax}
\providecommand{\BIBentryALTinterwordstretchfactor}{4}
\providecommand{\BIBentryALTinterwordspacing}{\spaceskip=\fontdimen2\font plus
\BIBentryALTinterwordstretchfactor\fontdimen3\font minus \fontdimen4\font\relax}
\providecommand{\BIBforeignlanguage}[2]{{%
\expandafter\ifx\csname l@#1\endcsname\relax
\typeout{** WARNING: IEEEtran.bst: No hyphenation pattern has been}%
\typeout{** loaded for the language `#1'. Using the pattern for}%
\typeout{** the default language instead.}%
\else
\language=\csname l@#1\endcsname
\fi
#2}}
\providecommand{\BIBdecl}{\relax}
\BIBdecl

\bibitem{cebeci2024scaling}
C.~Cebeci, O.~D. Noroozi, and U.~Madhow, ``Scaling mmwave {MU-MIMO}: Information-theoretic guidance using real-world data,'' in \emph{2024 58th Asilomar Conference on Signals, Systems, and Computers}.\hskip 1em plus 0.5em minus 0.4em\relax IEEE, 2024, pp. 1620--1624.

\bibitem{noroozi2025scaling}
O.~D. Noroozi, J.~Han, W.~Tang, Z.~Zhang, and U.~Madhow, ``Scaling wideband massive {MIMO} radar via beamspace dimension reduction,'' in \emph{2025 61st Allerton Conference on Communication, Control, and Computing Proceedings}.\hskip 1em plus 0.5em minus 0.4em\relax Allerton Conference on Communication, Control, and Computing, 2025.

\bibitem{gao2021wideband}
F.~Gao, B.~Wang, C.~Xing, J.~An, and G.~Y. Li, ``Wideband beamforming for hybrid massive {MIMO} terahertz communications,'' \emph{IEEE Journal on Selected Areas in Communications}, vol.~39, no.~6, pp. 1725--1740, 2021.

\bibitem{dai2022delay}
L.~Dai, J.~Tan, Z.~Chen, and H.~V. Poor, ``Delay-phase precoding for wideband thz massive {MIMO},'' \emph{IEEE Transactions on Wireless Communications}, vol.~21, no.~9, pp. 7271--7286, 2022.

\bibitem{fontenauQuadratic}
C.~Fonteneau, M.~Crussière, and B.~Jahan, ``A systematic beam broadening method for large phased arrays,'' in \emph{2021 Joint European Conference on Networks and Communications \& 6G Summit (EuCNC/6G Summit)}, 2021, pp. 7--12.

\bibitem{sayidmarie2013synthesis}
K.~H. Sayidmarie and Q.~H. Sultan, ``Synthesis of wide beam array patterns using quadratic-phase excitations,'' \emph{International Journal of Electromagnetics and Applications}, vol.~3, no.~6, pp. 127--135, 2013.

\bibitem{zhang2025beam}
M.~Zhang, Y.~Liu, S.~Zhu, and A.~Zhang, ``Beam broadening design for large-scale antenna arrays using convex quadratic programming,'' \emph{IEEE Transactions on Aerospace and Electronic Systems}, 2025.

\bibitem{petersson2022energy}
S.~O. Petersson and M.~A. Girnyk, ``Energy-efficient design of broad beams for massive {MIMO} systems,'' \emph{IEEE Transactions on Vehicular Technology}, vol.~71, no.~11, pp. 11\,772--11\,785, 2022.

\bibitem{sohrabi2017hybrid}
F.~Sohrabi and W.~Yu, ``Hybrid analog and digital beamforming for mmwave {OFDM} large-scale antenna arrays,'' \emph{IEEE Journal on Selected Areas in Communications}, vol.~35, no.~7, pp. 1432--1443, 2017.

\end{thebibliography}

\end{document}